\documentclass[11pt]{article}
\usepackage{moriond,epsfig}

\def\be{\begin{equation}}
\def\ee{\end{equation}}
\def\bea{\begin{eqnarray}}
\def\eea{\end{eqnarray}}

\begin{document}
\vspace*{4cm}
\title{Heavy-flavour production in pp and Pb--Pb collisions at ALICE}

\author{Andr\'e Mischke~\footnote{Email: a.mischke@uu.nl} for the ALICE Collaboration }

\address{ERC-Research Group {\em QGP-ALICE}, Institute for Subatomic Physics, \\
Utrecht University, Princetonplein 5, 3584 CC Utrecht, the Netherlands}

\maketitle

\abstracts{
In this contribution, recent ALICE measurements of the production cross section of single electrons, single muons and open charmed mesons in proton-proton collisions at $\sqrt{s}$~= 7 TeV are reported.
The data are compared to next-to-leading-order perturbative QCD calculations.
First open charm signals in Pb--Pb collisions at $\sqrt{s_{\rm NN}}$~= 2.76~TeV are shown.}

\section{Introduction}
Heavy quarks (charm and beauty) are sensitive penetrating probes to study the properties of the hot quark matter state that is formed in collisions of heavy atomic nuclei at high energy densities. 
Due to their large mass, they are believed to be predominantly produced in the early stage of the collision by 
gluon-fusion processes, so that they explore the entire evolution of the produced matter. 
The Large Hadron Collider (LHC) at CERN allows to create and carefully study such matter in heavy-ion collisions at an unprecedented energy. 
ALICE collected a rich pp data sample at a collision energy of 7 TeV.
These measurements are important to test predictions from pQCD (such as the D meson production cross sections) in the new energy domain and provide an essential baseline for the comprehensive studies in heavy-ion collisions.

\section{ALICE detector, trigger and data set}
\begin{figure}[t]
  \centering
  \hspace{2mm}
  \includegraphics[scale=0.29]{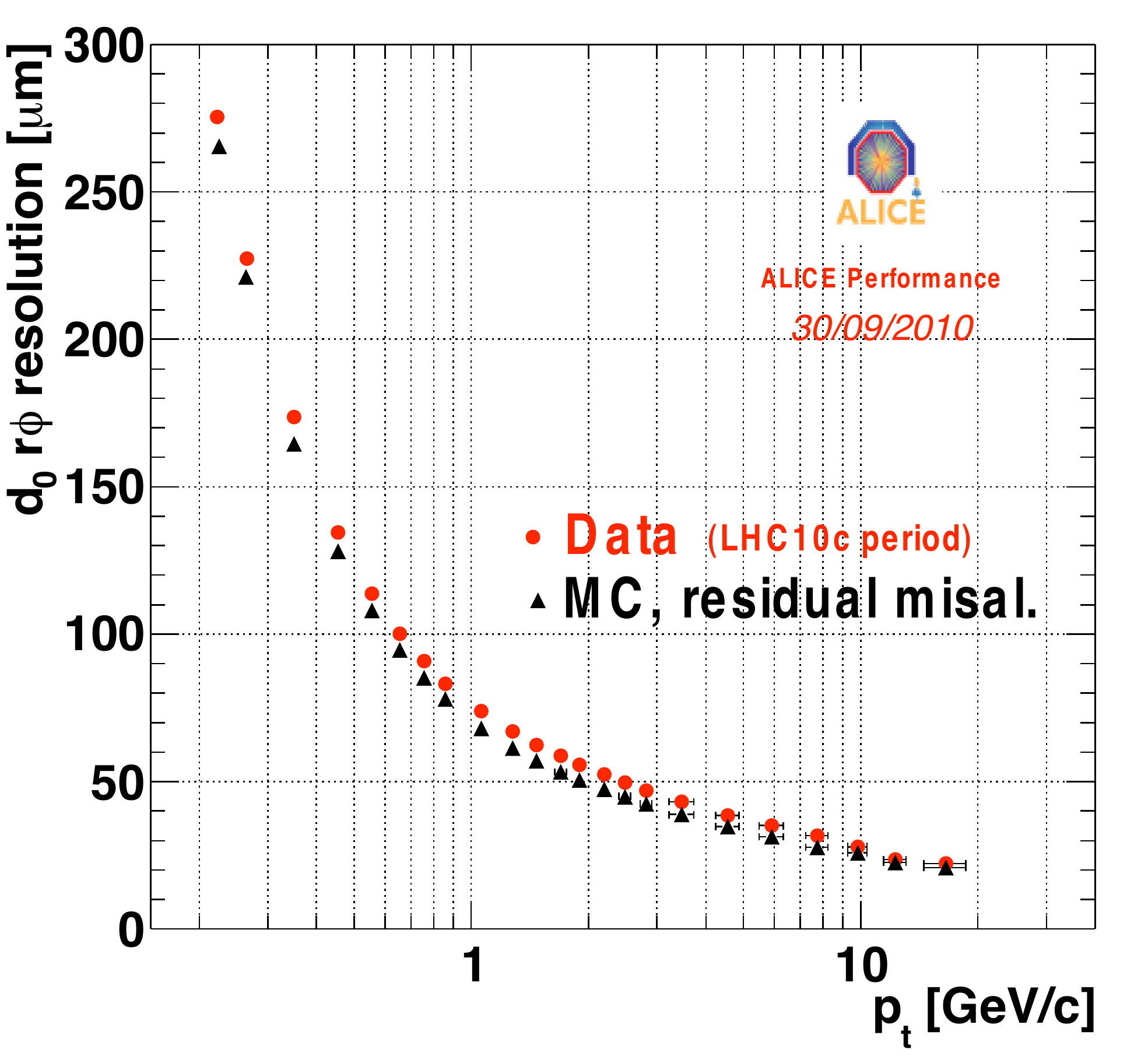}
  \hspace{1mm}
  \includegraphics[scale=0.35]{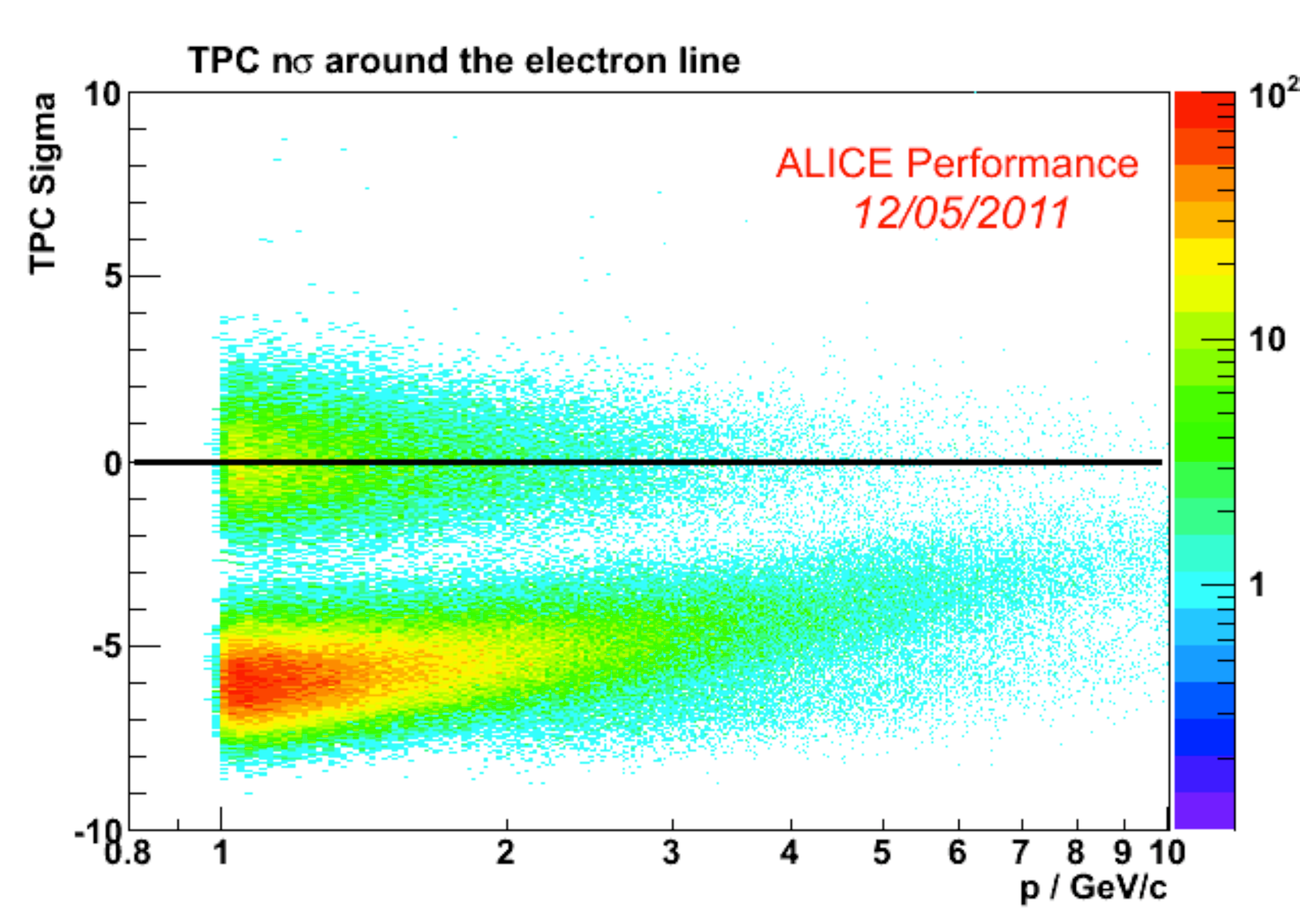}
  \vspace*{-0.4cm}
  \caption{Left: $p_{\rm T}$ dependence of the track impact parameter resolution in the transverse plane ($r$ direction) for data and Monte-Carlo simulations. Right: Electron identification using the normalized TPC d$E$/d$x$ distribution. The upper band are electrons and the lower ones charged pions.}  \label{Fig:1}
\end{figure}

ALICE, A Large Ion Collider Experiment, is the dedicated detector for measurements in heavy-ion collisions~\cite{alice:PPR2}~\cite{alice:detpaper}, demonstrated in the first LHC Pb--Pb run~\cite{alice:first}.
Its characteristic features are the very low momentum cut-off (100 MeV/$c$), the low material budget and the excellent particle identification and vertexing capabilities in a high multiplicity environment.
Tracking and particle identification through the measurement of the specific energy loss (d$E$/d$x$) is performed using the large volume Time Projection Chamber (TPC) located inside the large L3 solenoidal magnet with a field of $B$ = 0.5 T.
The TPC has a coverage of -0.9 to 0.9 in pseudo-rapidity and 2$\pi$ in azimuth.
%
The particle identification at about $p_{\rm T} >$ 2 GeV/$c$ is performed with the Time Of Flight (TOF) detector.
%
The Inner Tracking System consists of six concentric cylindrical layers of silicon detectors and provides excellent reconstruction of displaced vertices with a transverse impact parameter resolution better than 75 $\mu$m for $p_{\rm T} >$ 1 GeV/$c$ (cf. Fig.~\ref{Fig:1} (left panel)).
The minimum bias event selection is based on a signal in either of the VZERO scintillator counters or at least one hit in one of the two innermost silicon layers.
The results presented in this paper correspond to 100M and 180M minimum-bias pp events at $\sqrt{s}$ = 7 TeV for the D meson (integrated luminosity: 1.6~nb$^{-1}$) and single electron analyses (2.6~nb$^{-1}$), respectively, and 2M muon triggered events (3.49~nb$^{-1}$).
%

\section{D mesons}
\begin{figure}[t]
  \centering
  \includegraphics[scale=0.26]{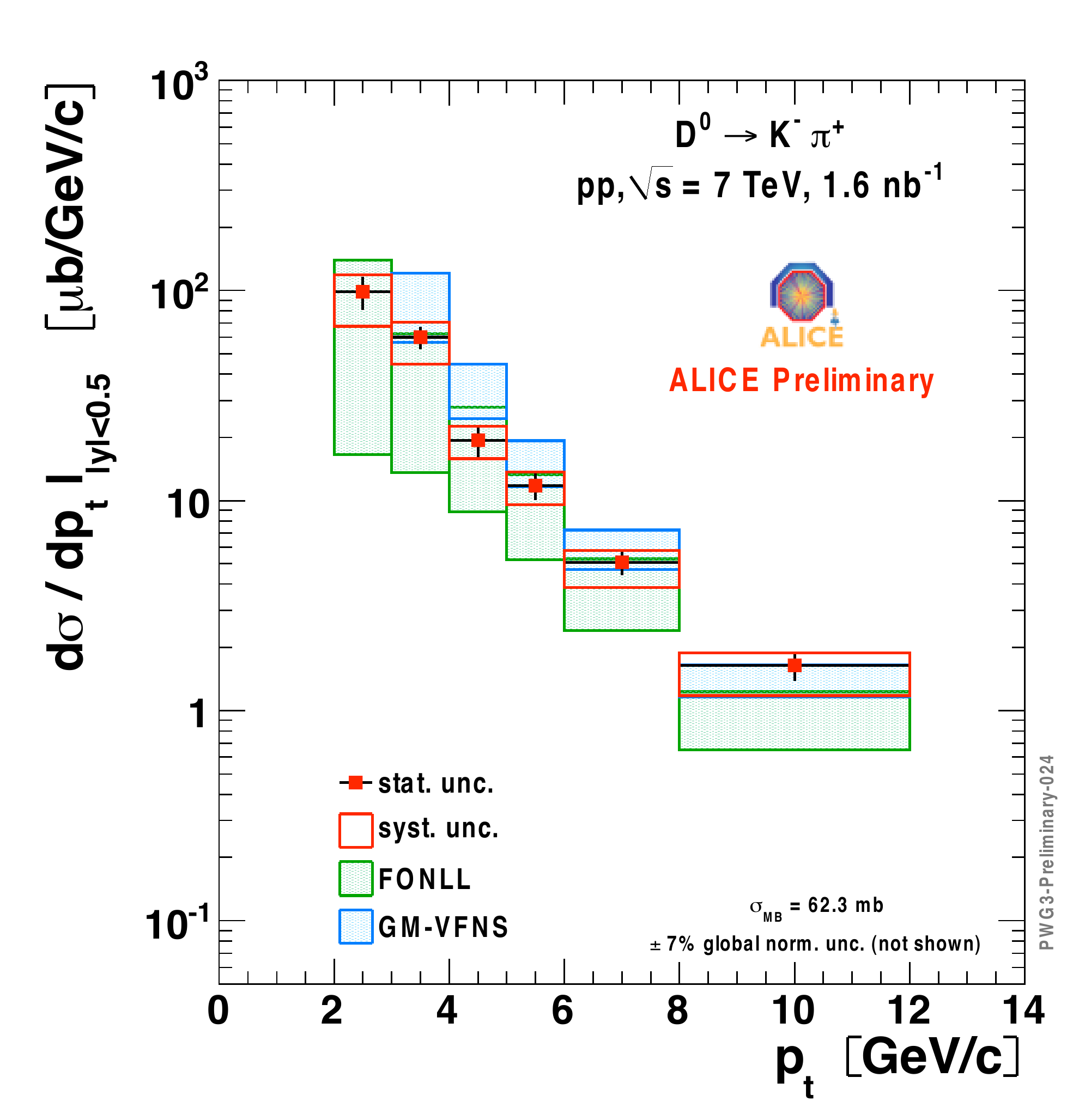}
  \includegraphics[scale=0.26]{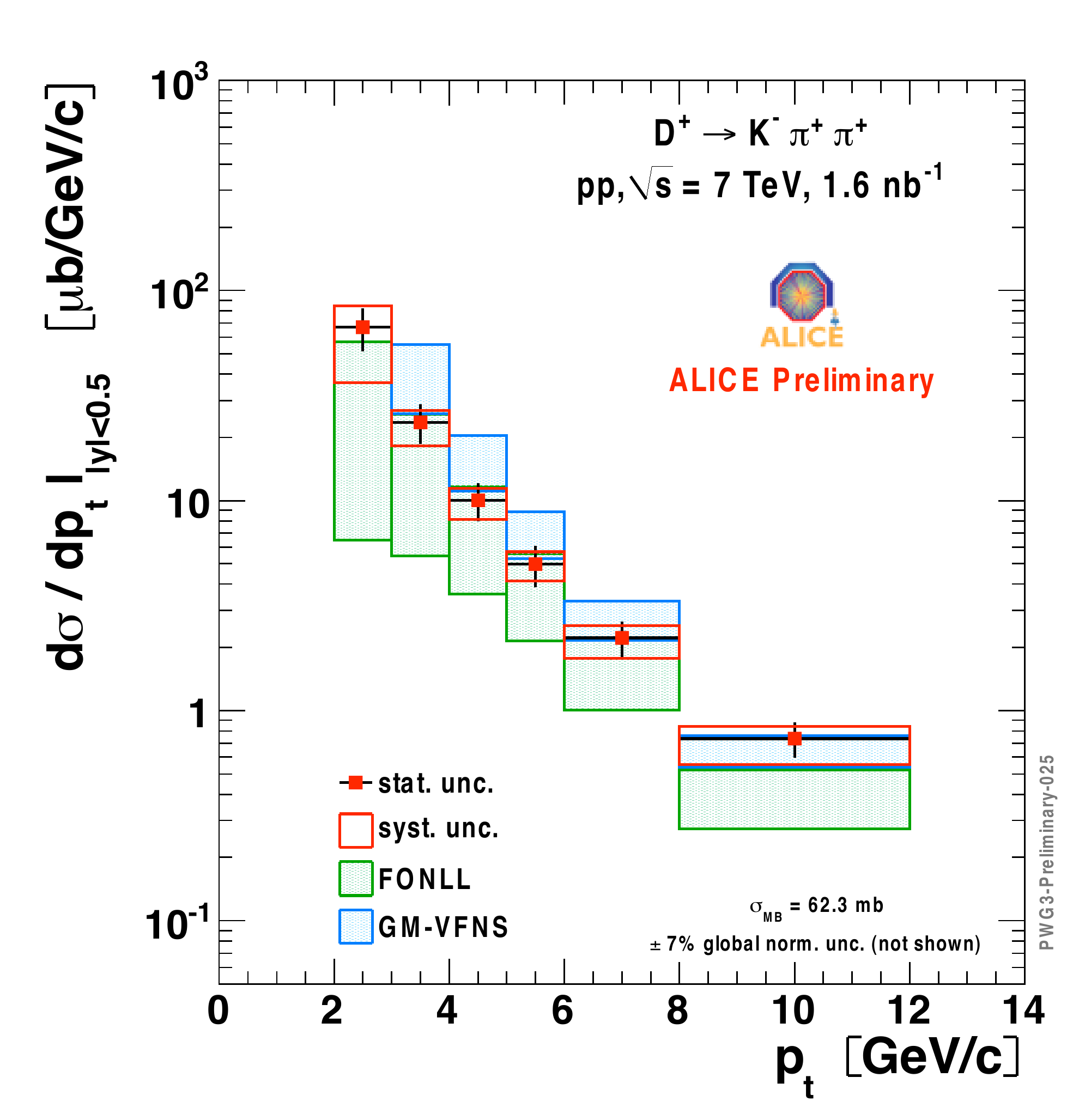}
  \includegraphics[scale=0.26]{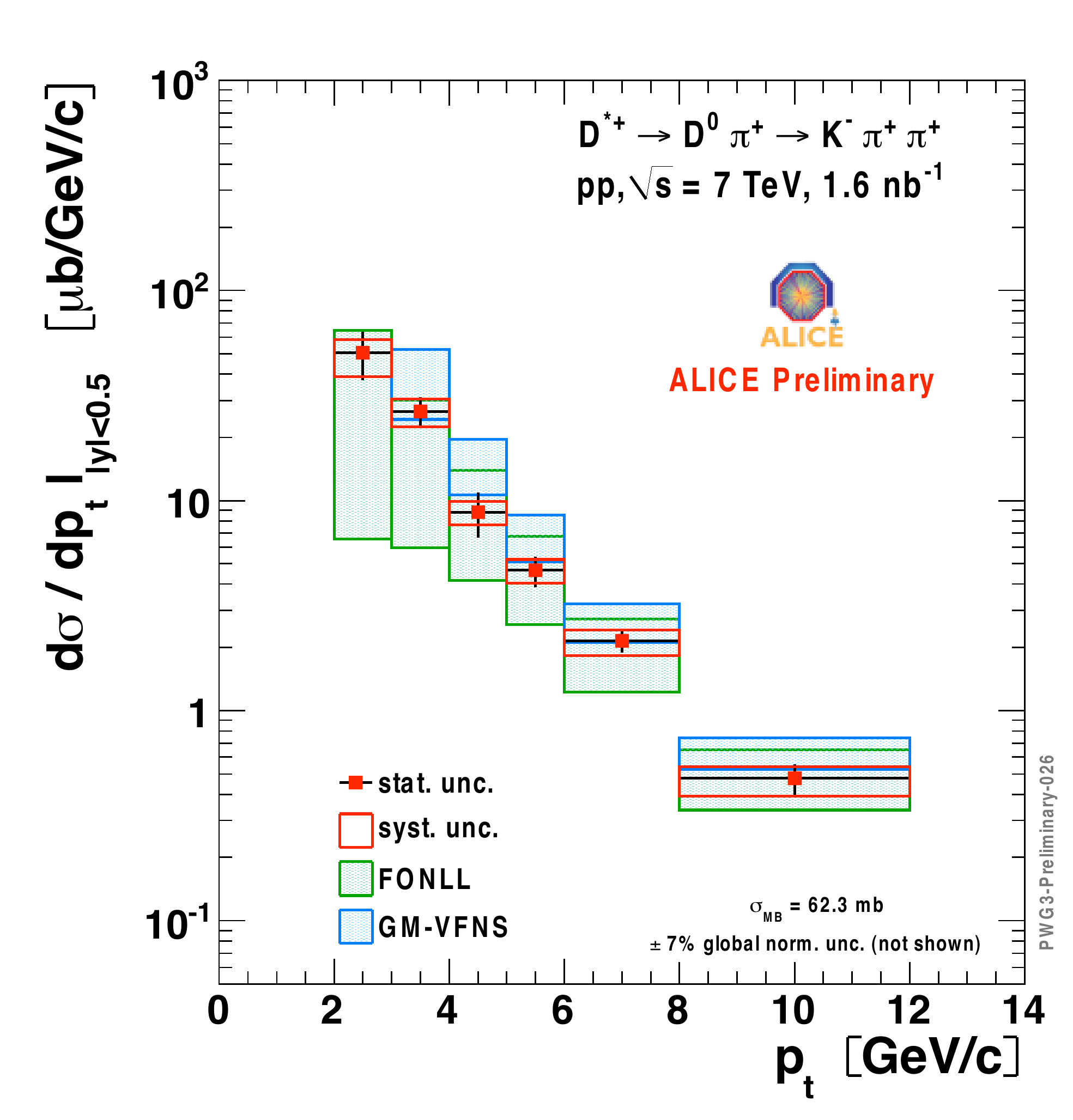}
  \vspace*{-0.3cm}
  \caption{$p_{\rm T}$ differential cross section for D$^0$, D$^+$ and D$^{*+}$ mesons in pp collisions at $\sqrt{s}$~= 7 TeV, compared to next-to-leading-order perturbative QCD predictions from FONLL and GM-VFNS calculations.}  \label{Fig:2}
\end{figure}

\begin{figure}[t]
  \centering
  \includegraphics[scale=0.17]{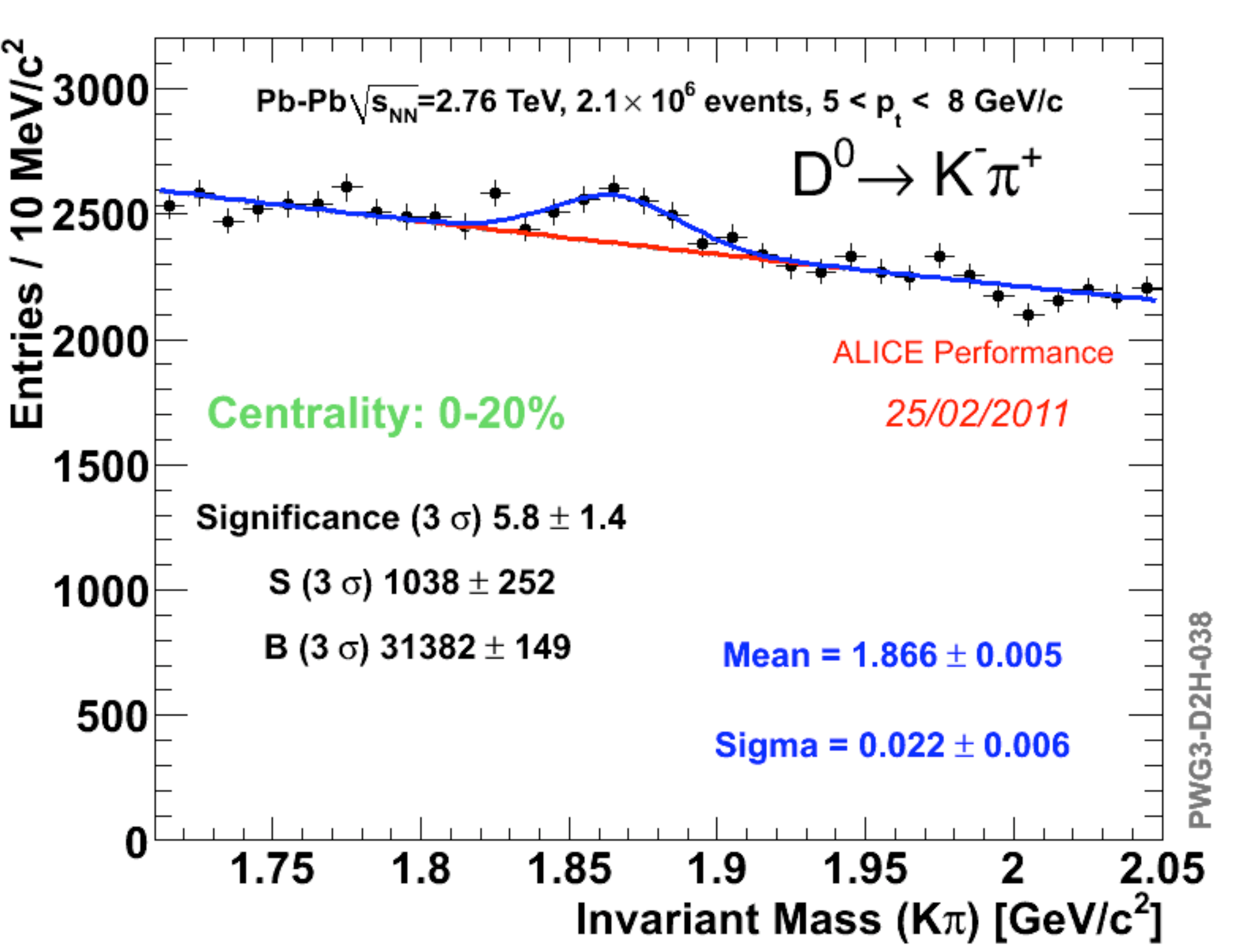}
  \includegraphics[scale=0.15]{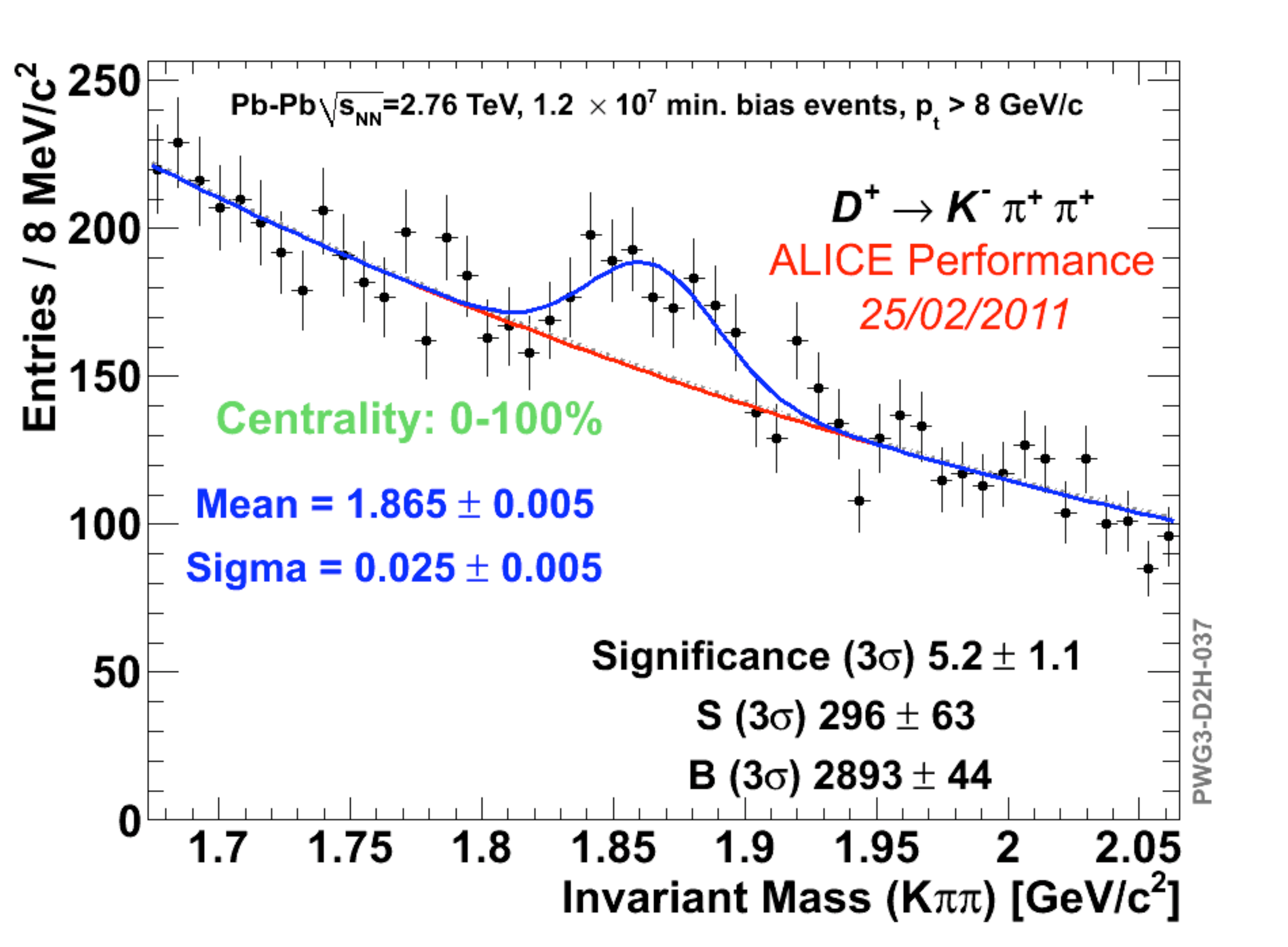}
  \includegraphics[scale=0.23]{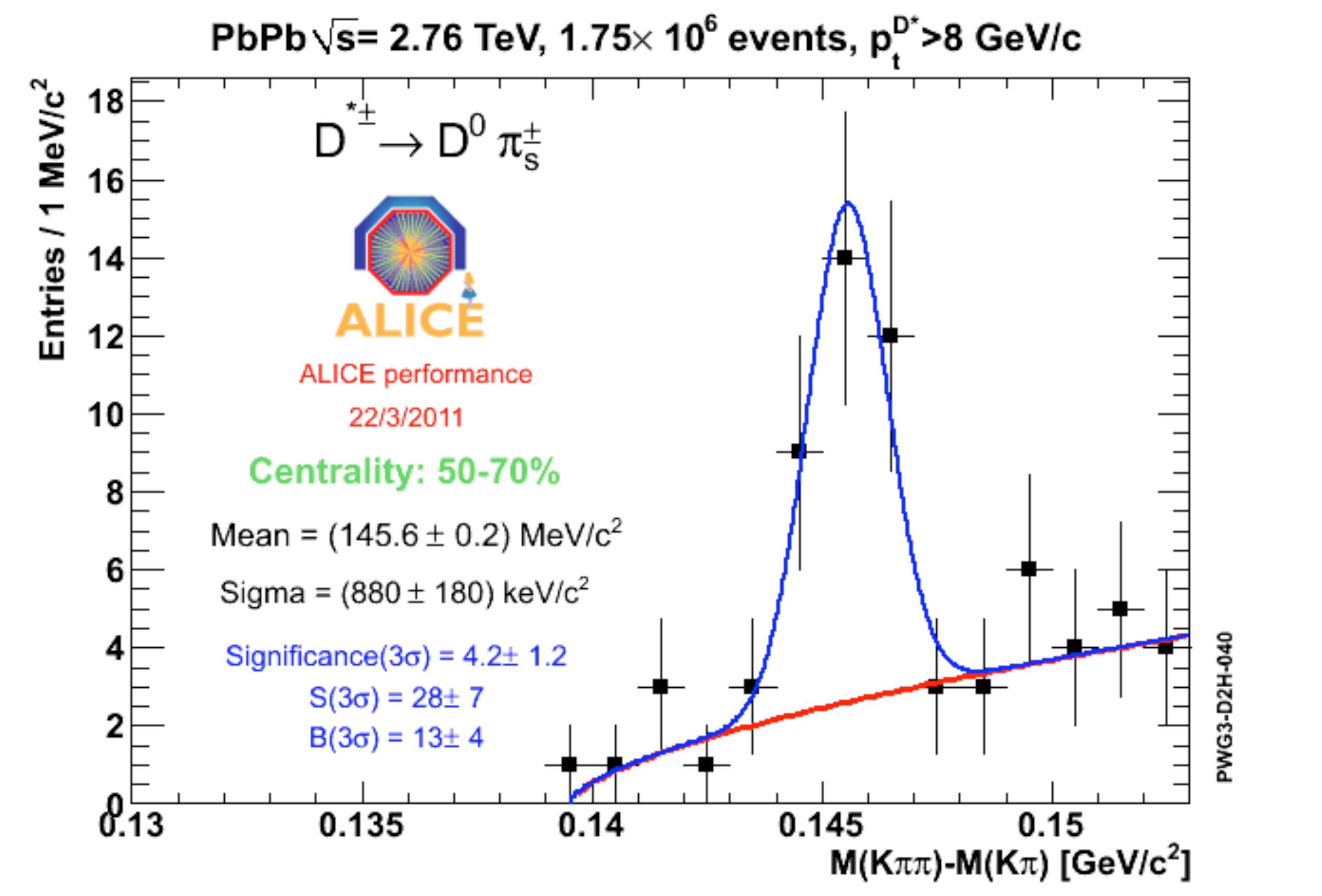}
  \vspace*{-0.2cm}
  \caption{Invariant mass distribution for $K^-\pi^+$ (left panel), $K^-\pi^+\pi^+$ pairs (middle panel) and $M(K\pi\pi)-M(K\pi)$ (right panel) in lead-lead collisions at 2.76 TeV per nucleon-nucleon pair.}  \label{Fig:3}
\end{figure}

The reconstruction of D mesons is based on their decay topology and the invariant mass technique. Details of the analysis can be found in~\cite{d:grelli}.
The kaon and pion identification using TPC and TOF helps to reduce background at low $p_{\rm T}$.
Figure~\ref{Fig:2} shows the $p_{\rm T}$ differential cross section for prompt D$^0$, D$^+$ and D$^{*+}$ mesons using 20\% of the 2010 statistics.
The feed-down from beauty decays is calculated from theory and gives a contribution of 10-15\%. A data driven method will be used with the full 2010 statistics.
The data are well described by pQCD calculations at Fixed-Order plus Next-to-Leading Logarithm (FONLL) level~\cite{theo:fonll} and GM-VFNS predictions~\cite{theo:vfns}.
The open charm signals in the first Pb--Pb collisions at $\sqrt{s_{\rm NN}}$~= 2.76~TeV (cf. Fig.~\ref{Fig:3}) indicate that a detailed study of the modification of the D meson production in hot quark matter will be possible with the 2010 data sample.
%

\section{Single electrons}
\begin{figure}[t]
  \centering
  \hspace{2mm}
  \includegraphics[scale=0.24]{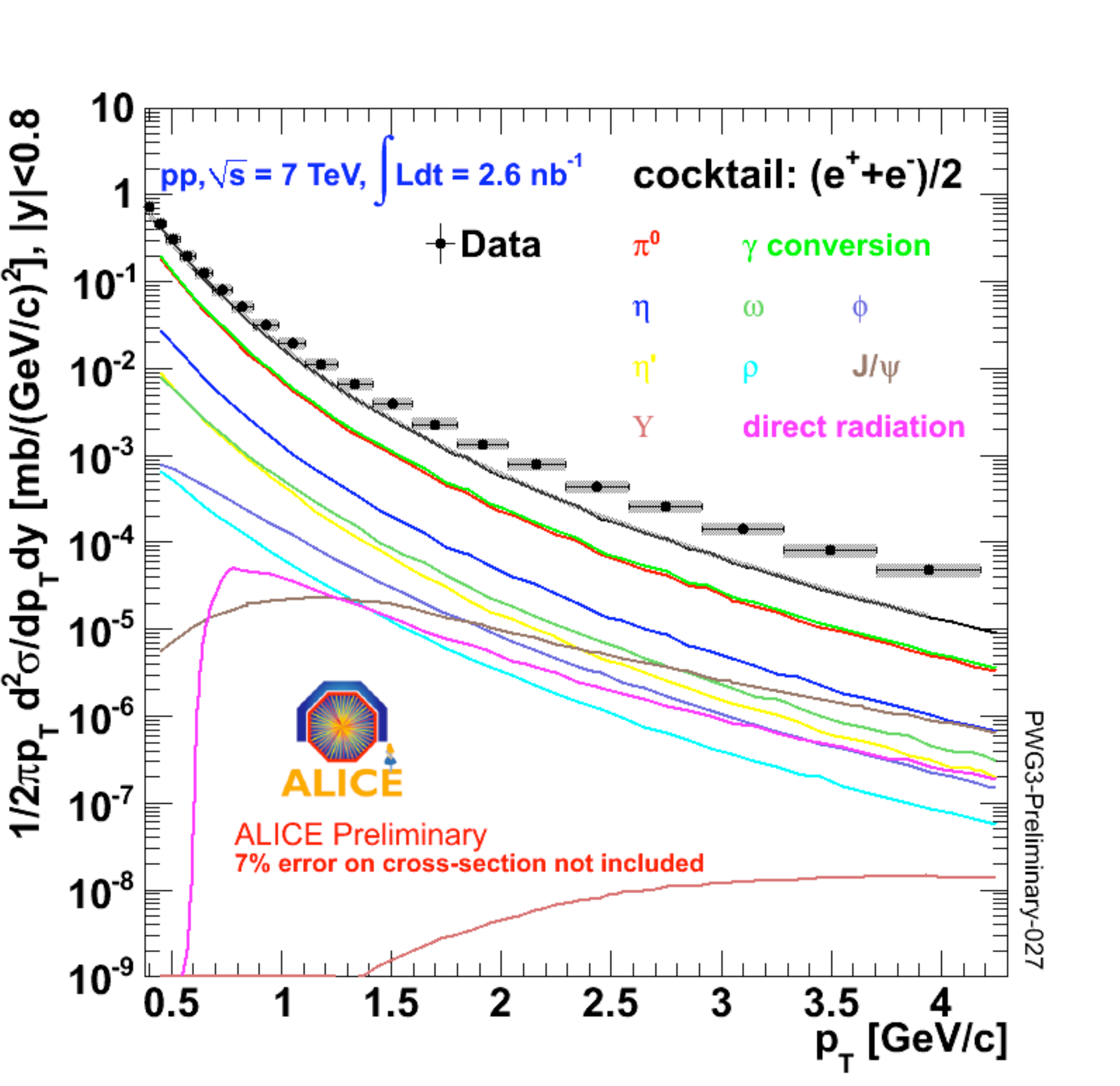}
  \hspace{1mm}
  \includegraphics[scale=0.24]{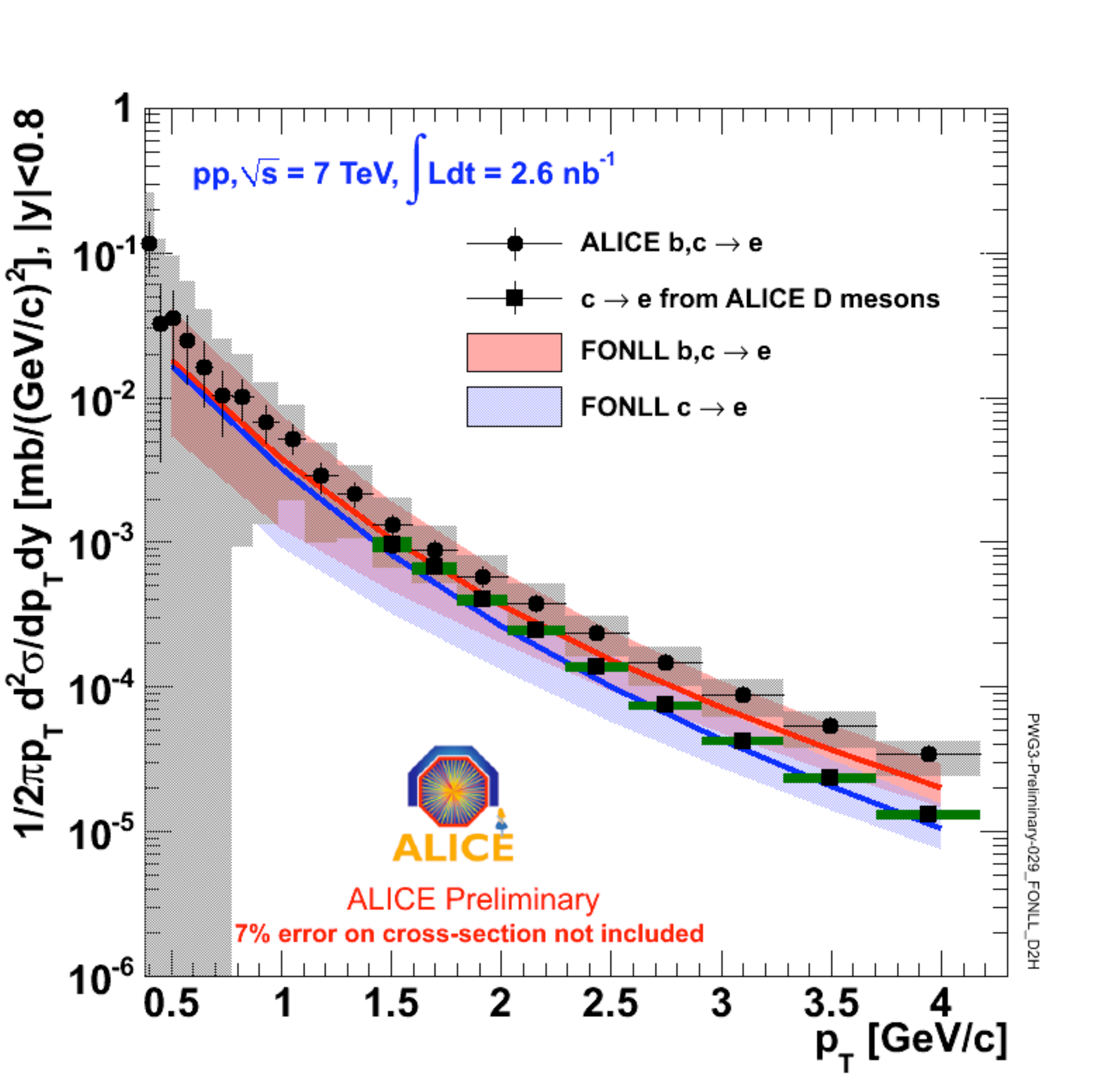}
  \vspace*{-0.3cm}
  \caption{Invariant inclusive electron cross section, compared to the cocktail of background electrons (left panel) and invariant single electron cross section, compared to FONLL calculations and to expectations from D meson decay electrons (right panel).}  \label{Fig:4}
\end{figure}

The open heavy-flavour production cross section at mid-rapidity has also be studied through the measurement of single electrons. 
Those electrons are identified on a statistical basis by subtracting a cocktail of background electrons from the inclusive electron spectrum. 
This background arises mainly from electrons from $\gamma$ conversion in the detector material and $\pi^0$ Dalitz decays. 
For $p_{\rm T}$ up to a few GeV/$c$ this cocktail can be determined precisely by means of the measured $\pi^0$ cross section.
Figure~\ref{Fig:1} (right panel) depicts the momentum dependence of the normalized TPC d$E$/d$x$ distribution after applying a cut on the TOF signal, which rejects kaons ($<$ 1.5 GeV/$c$) and protons ($<$ 3 GeV/$c$).
A d$E$/d$x$ cut clearly separates electrons from charged pions up to about 10 GeV/$c$ with a residual pion contamination of less than 15\%.
The corrected inclusive electron spectrum is shown in Fig.~\ref{Fig:4} (left panel) together with the cocktail of background electrons.
Figure~\ref{Fig:4} (right panel) illustrates the single electron cross section, which has a total systematic uncertainty of 16-20\% ($p_{\rm T}$ dependent) plus 7\% for the normalization.
The data are well described by FONLL calculations~\cite{theo:fonll} within errors. 
Moreover, the low $p_{\rm T}$ single electron spectrum agrees with expectations from D meson decay electrons.
The $p_{\rm T}$ range will be extended with the TRD and EMCAL detectors in the near future.
Electrons from beauty decays will be identified through displaced vertices.
%

\section{Single muons}
\begin{figure}[t]
  \centering
  \includegraphics[scale=0.29]{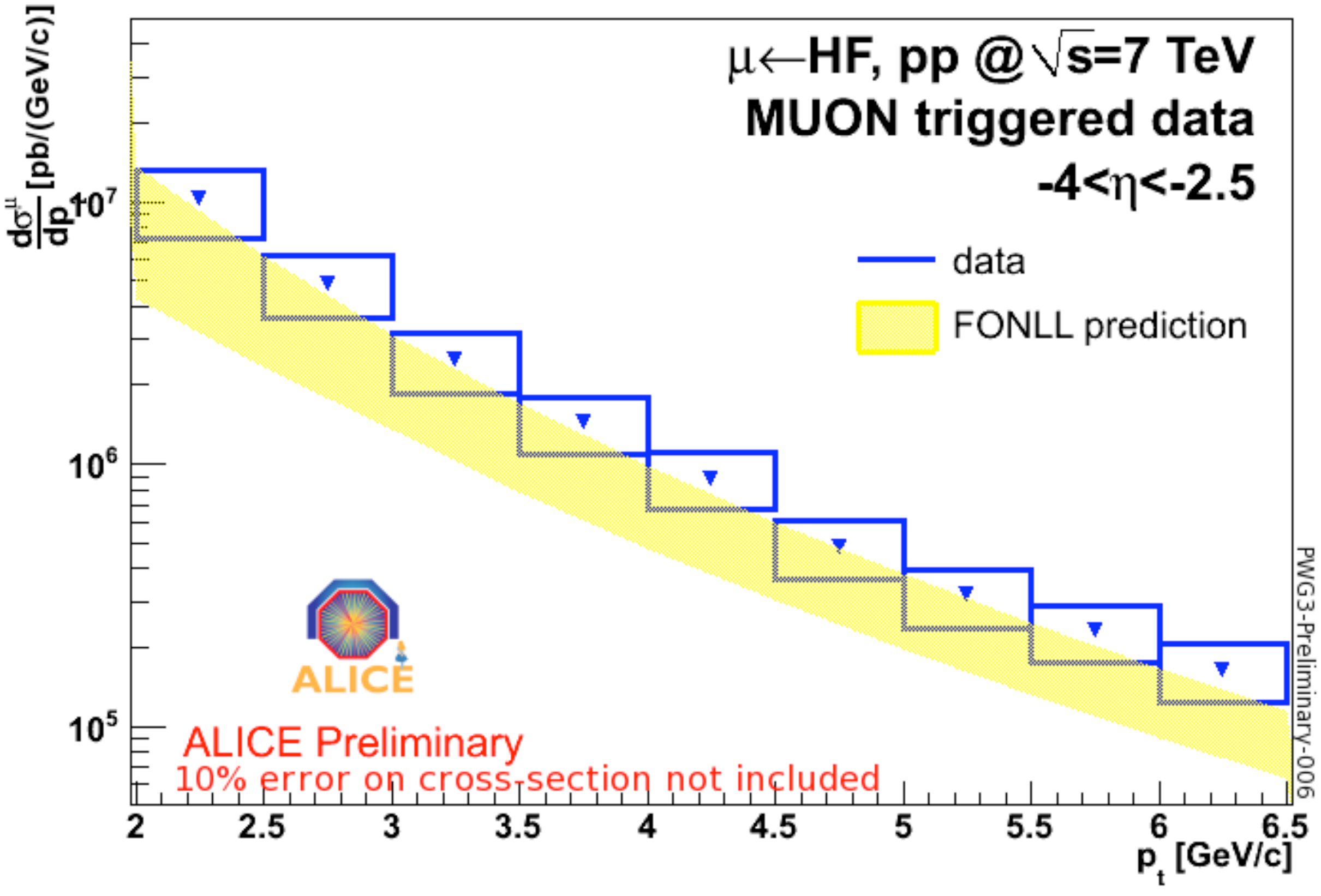}
  \vspace*{-0.3cm}
  \caption{Differential transverse momentum cross section for single muons in $-4<\eta<-2.5$. The statistical error is smaller than the markers. The systematic errors (open boxes) do not include an additional 10\% error on the minimum-bias pp cross section. The yellow band indicates the FONLL prediction.}  \label{Fig:5}
\end{figure}

Heavy-flavour production at forward rapidities can be studied in ALICE with single muons using the muon spectrometer, which covers an $\eta$-range from -4 to -2.5.
The extraction of the heavy-flavour contribution of the single muon spectra requires the subtraction of three main background sources: a) muons from the decay-in-flight of light hadrons (decay muons); 
b) muons from the decay of hadrons produced in the interaction with the front absorber (secondary muons); c) punch-through hadrons.
The last contribution can be efficiently rejected by requiring the matching of the reconstructed tracks with the tracks in the trigger system. 
Due to the lower mass of the parent particles, the background muons have a softer $p_{\rm T}$ distribution than the heavy-flavour muons, and dominate the low-$p_{\rm T}$ region.  Therefore, the analysis is restricted to the $p_{\rm T}$ range 2-6.5 GeV/$c$, where the upper limit is being determined by the current $p_{\rm T}$ resolution of spectrometer with partial alignment. 
Simulation studies indicate that, in this $p_{\rm T}$ range, the contribution of secondary muons is small (about 3\%). 
The main source of background in this region consists of decay muons (about 25\%), which have been subtracted using Monte Carlo simulations. 
Figure~\ref{Fig:5} illustrates the single muon $p_{\rm T}$ cross section after corrections. The systematic uncertainties are 20-25\%.
The FONLL calculation~\cite{theo:fonll} agrees with the data within uncertainties. 
%
%

\section{Summary}
Recent ALICE results on open charmed mesons and single leptons in pp collisions at 7~TeV are presented.
The production cross-section of single electrons, single muons and D mesons are measured up to $p_{\rm T}$~= 4, 6.5 and 12 GeV/$c$, respectively, and show good agreement with NLO pQCD calculations.
First D meson signals are shown for Pb--Pb collisions at $\sqrt{s_{\rm NN}}$~= 2.76~TeV.

\section*{Acknowledgments}
The European Research Council has provided financial support under the European Community's Seventh Framework Programme (FP7/2007-2013) / ERC grant agreement no 210223.
This work was supported in part by a Vidi grant from the Netherlands Organisation for Scientific Research (project number 680-47-232).

\section*{References}

\end{document}